\documentclass[aps,prl,twocolumn,showpacs,superscriptaddress,groupedaddress]{revtex4} 
\usepackage{graphicx}
\usepackage{amssymb} 
\usepackage{amsmath}
\usepackage{braket}
\begin{document}

\title{Possible Astrophysical Observables of Quantum Gravity Effects near Black Holes
}

\author{Ue-Li Pen
}
\affiliation{
Canadian Institute for Theoretical Astrophysics, Toronto,
  Ontario, Canada\\
}

\date{\today}

\begin{abstract}

Recent implications of results from quantum information theory applied
to black holes has led to the confusing conclusions that requires
either abandoning the equivalence principle (e.g. the firewall
picture), or the no-hair theorem (e.g. the fuzzball picture), or even
more impalatable options.

The recent discovery of a pulsar orbiting a black hole opens up new
possibilities for tests of theories of gravity.  We examine possible
observational effects of semiclassical quantum gravity in the vicinity
of black holes, as probed by pulsars and event horizon telescope
imaging of flares.  Pulsar radiation is observable at wavelengths only
two orders of magnitude shorter than the Hawking radiation, so
precision interferometry of lensed pulsar images may shed light on the
quantum gravitational processes and interaction of Hawking radiation
with the spacetime near the black hole.  This paper discusses the
impact on the pulsar radiation interference pattern, which is
observable through the modulation index in the foreseeable future, and
discusses a possible classical limit of BHC.

\end{abstract}                        
\maketitle 

\newcommand{\be}{\begin{eqnarray}}
\newcommand{\ee}{\end{eqnarray}}
\newcommand{\beq}{\begin{equation}}
\newcommand{\eeq}{\end{equation}}

\textit{Introduction --} The recent discovery of
PSRJ1745-2900\cite{2013MNRAS.435L..29S,2013ApJ...775L..34R,2013Natur.501..391E}
orbiting the galactic center black hole opens up new possibilities for
precision tests of gravity.  It allows us to investigate possible
outcomes as its orbit is mapped, and possible quantum deviations from
standard Einstein gravity.

It has proven challenging to find experimental consequences of quantum
gravitational effects.  At the same time, precision experimental
probes of classical general relativity have a dearth of alternate
theories to compare with.

In this letter, we explore possible semi-classical consequences of
pulsar-black hole binaries and flares in the galactic center black
hole accretion flow.  This is meant to stimulate concrete discussions
of quantum mechanics applied to gravitational systems in scenarios
that may be testable in the foreseeable future.

{\it Motivation} -- The quantum mechanical nature of black holes has
provided a fruitful testbed for thought experiments and discussions.
Hawking's calculation led to the possibility of black hole radiation
and evaporation.  The radiation appears thermal, and appears not to
depend on the interior of the black hole, or its formation history.
This leads to the well known information loss
problem\cite{1976PhRvD..14.2460H}.

Historically the resolution of the problem included violation of
unitarity (i.e. causality), or the possibility of remnants.  String
theory is a constructive example which is unitary and contains the
same black hole entropy and evaporation, and no remnants.  In this
context, the resolution of the paradox has to lie in the purity of
Hawking radiation vs the breakdown of the equivalence principle near
the horizon.  As discussed in \cite{2013JHEP...02..062A} (hereafter
AMPS), a modification of Hawking radiation purity requires macroscopic
changes in space-time of order unity outside the Scharzschild radius
(see also \cite{,2009CQGra..26v4001M} for a complementary view).
Different groups arrive at opposite aesthetic conclusions from this
line of reasoning: the firewall\cite{2013JHEP...02..062A} picture
maintains radiation purity, and instead sacrifices the equivalence
principle for infalling observers, who burn up at the horizon, thus
preventing them from measuring violations of quantum mechanics.  The
fuzzball picture explores the opposite path\cite{2013JHEP...09..012A}:
the Hawking photons are emitted by a substantially non-Scharzschild
geometry, but classical observers see a general relativistic spacetime
including the equivalence principle on both sides of the horizon.
This latter framework is consistent with principle of Black Hole
Complementarity\cite{1993PhRvD..48.3743S} (BHC, see also the Fuzzball
interpretation of BHC\cite{2013JHEP...09..012A}), that the classical
and quantum pictures depend on the nature of the measurement.
Spefically, low energy probes, such as Hawking radiation or grazing
pulsar radiation, would be subject to the quantum nature, while high
energy probes, including protons, stars, and other likely matter see a
classical space-time.  In this \textit{Letter} we follow this scenario
to explore possible consequences for quantum measurement using pulsar
radiation, which provide realistic probes of BHC, using wavelengths
comparable in energy to the Hawking radiation.  This opens up the
possibility of testing physics using real experiments instead of
aesthetic considerations.

Pulsars are highly compact, very bright light sources and exquisite
clocks, enabling precision measurements of space-time.  A pulsar
orbiting a black hole provides a scenario which accentuates potential
experimental outcomes.  Very recently, the first candidate has been
discovered
\cite{2013MNRAS.435L..29S,2013ApJ...775L..34R,2013Natur.501..391E},
likely orbiting the galactic center black hole.  While the orbital
parameters are not known to the author, there is a possibility that
its orbit in projection passes close behind the black hole, such that
a gravitationally lensed image becomes visible. Depending on orbital
parameters, such a conjunction could take decades to occur during
which time more pulsar-black hole binaries may be discovered.  In
addition to the galactic centre supermassive black-hole pulsar binary,
ten double neutron star systems are known, and discovery of a
pulsar-stellar mass black hole binary appears
likely\cite{2002ApJ...572..407B}.  For purposes of this discussion,
slow and fast pulsars are both suitable, with slow pulsars dominating
the predicted populations. This is one of the goals of the planned
Square Kilometer Array\footnote{www.skatelescope.org}.

{\it Strong Gravitational Lensing --} We consider the dynamics of a
pulsar orbiting a black hole.  As a pulsar passes behind a black hole,
multiple images of the pulsar appear.  In the weak field limit, one
sees two images. This phenomenon is called ``strong lensing''.  In the
strong field regime, an infinite number of exponentially fainter
images appear \cite{2011arXiv1110.2789B,2006PhRvD..73f3003R}.  In this
section, we will confine our discussion to the two images under ``weak
field'' strong gravitational lensing.

We will consider the regime where the pulsar is many
Schwarzschild radii behind the black hole, and the weak field limit
applies, with only small perturbations.  The pulsar radiation is
lensed by the black hole's gravitational field, which is well
described by geometric optics.  We first review the geometric optics,
and then estimate the interference pattern of this double-slit
experiment.

Generally, the brighter image is further away from the black hole, and less
affected by post Einsteinian effects.  This two image scenario is much
like a quantum double slit experiment.  The wavelength of the photons
is not drastically different from that of the thermally emitted
Hawking radiation photons, and is expected to probe the low energy limit.

A
background source always has two images: one inside the Einstein
radius, which we call the interior image, and one outside the Einstein
radius, the exterior image.
The Einstein radius is defined in \cite{1992grle.book.....S}:
\be
\theta_E=\frac{\sqrt{r_s D_{ds}}}{D_d},
\ee
where the Scharzschild radius $r_s = 2GM/c^2$, $D_{ds}$ is the distance from the black hole to
the pulsar, and $D_d$ is the distance to us. We are considering the
limit where $r_s \ll D_{ds} \ll D_d$. From here onward, we will
use units where the speed of light $c=1$.
The apparent image positions
for a source at angular separation  $\beta$
are at
\be
\theta_\pm=\frac{1}{2}\left(\beta \pm \sqrt{\beta^2+4\theta_E^2}  \right)
\ee
with magnifications
\be
\mu_\pm = \frac{u^2+2}{2 u\sqrt{u^2+4}}\pm \frac{1}{2}
\ee
where $u=\beta/\theta_E$\cite{1992ApJ...389L..41M}.  For large
separations, the interior image 
gets faint as $1/u^4$, while the exterior image goes to its unlensed
brightness.
The time delay between the two images is
\be
\Delta t=r_s\left[\frac{1}{2}u\sqrt{u^2+4}+\ln \frac{\sqrt{u^2+4}+u}{\sqrt{u^2+4}-u}\right]
\label{eqn:dt}
\ee

We now consider potential fuzzball corrections to this standard picture.  In
Scharzschild coordinates, the metric is
\be
ds^2=-[1-\psi(r)] dt^2 +[1-\psi(r)]^{-1} dr^2 + r^2 d\Omega
\ee
with $\psi(r)=r_s/r$.
For a light source many $r_s$ away, the lensing equation depends on
the projected potential $\varphi(\theta)=2\frac{D_{ds}}{D_d^2}\int
\psi(\sqrt{(\theta D_d)^2+z^2}) dz$.  We consider a general multipole
expansion of the potential, which in projection becomes
\begin{equation}
\varphi(\theta,\phi)=\theta_E^2\ln(\theta/\theta_E)+\theta_E^2\sum_m \frac{a_m \theta_s^m \cos[m (\phi-\phi_m)]}{\theta^m},
\end{equation}
with the apparent Schwarzschild radius $\theta_s$=$r_s/D_d$.  
We define the ratio $b\equiv \theta_E/\theta_s$, which is the impact
parameter of the lensed image in units of Schwarzschild radii.
The
generic orbit has $b \gg 1$, and our data probes $\theta
\sim \theta_E$. 
At these large radii, the low $m$ harmonics  dominate.  Since the
deviations to the space-time are of order unity near the horizon, one
expects $a_m$ to be order unity, and by isotropy the $\phi_m$ are
uniformly distributed.
In a firewall picture, all coefficients $a_m=0$.

The lowest order perturbation is a dipole, $m=1$.  This is analogous
to a displacement of the black hole position by a Schwarzschild
radius.  This results in a variation of time delay
\beq
\delta \Delta t = \alpha \frac{r_s}{b}
\label{eqn:sdelay}
\eeq
for a order unity proportionality constant $\alpha$.

In a fuzzball scenario, the delayed pulse will appear broadened: its
pulse profile will appear wider, convolved by the distribution of
delays from the different multipole perturbations.  While the actual
separation of images may be challenging to resolve in angle, the
delays are readily observable.  As in interstellar plasma lensing, the
multiple images interfere constructively and destructively when
observed in a single dish telescope\cite{2012MNRAS.421L.132P}.  For
time estimates, we will scale to a solar mass black hole.  When
applied to the galactic center supermassive black hole, we will
explicitly state that.  Typical pulsars have the most sensitive
detections at $\sim$ GHz.  In a classical black hole, this results in
a very precise measurement of delay and dopplershift: with a 1 GHz
bandwidth, delays are measurable to a nanosecond.  The characteristic
delay $\Delta t$ for a solar mass black hole is several microseconds,
and expected fuzzball fluctuations smaller by $b$.  For a given set of
orbital parameters, the fringe pattern is fully determined, and can be
tracked.  After shifting the self-interference pattern by the
geometric delay, one is sensitive to the stochastic quantum
fluctuations in time delay (\ref{eqn:sdelay}), which induce phase
differences of order unity for $b \lesssim 1000$.

\begin{figure}
\begin{center}
\includegraphics[width=0.48\textwidth]{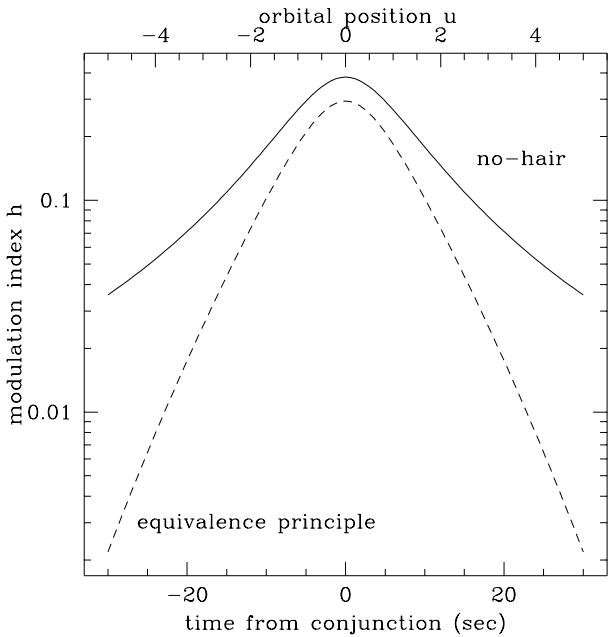}
\end{center}
\vspace{-0.7cm}
\caption{\label{fig:mod}
Pulsar modulation index.  The aesthetic choice of ``no-hair'' is
thought to result in a
firewall, with classical modulation index shown by the solid line.
The alternative choice embracing a  equivalence/complementarity principle,
and thus rejecting the firewall, could lead to the dashed line, where
the modulation index is reduced due to exterior quantum effects.
}
\end{figure}

To estimate the observational signature, we consider a pulsar orbiting
a 10 M$_\odot$ BH, at $4\times 10^4 r_S$.  The orbital period is 2
hours, and gravitational radiation decay time $\sim 10^7$ yr,
comparable to the life time of a slow pulsar (the most common
type\cite{2005ApJ...628..343P}).  We use an inclination of a quarter
degree, which leads to conjunction at one Einstein radius.  The
modulation index $h$\cite{2012hpa..book.....L} resulting from the
interference pattern depends on the orbital phase, and is shown in
Figure \ref{fig:mod}.  We model the phase change in (\ref{eqn:sdelay})
as a Gaussian random field, which changes every dynamical time.  For
illustration convenience, we took the prefactor $\alpha\sim 0.4$.  In
a future data set, one could fit for $\alpha$ from the data.  The
first principles computation of $\alpha$ is beyond the scope of this
Letter, which we leave as a future exercise.  We used an observing
wavelength of $\lambda =20$m, the low end of the
LOFAR\footnote{http://www.lofar.org/} telescope. If the equivalence
principle is obeyed, the modulation index is decreased, as is apparent
in this picture.  The fractional effect is smallest at superior
conjunction, when the inner image is the furthest from the black hole,
and thus has the least quantum effects.

%By modeling the interference between the two images, and comparing
%with the uniquely predicted einstein interference, one could probe
%much finer differences.  Orbits of lesser inclination still show
%effects, but requires a larger sample to find one sufficiently
%inclined.  The actual requirement on inclination is a function of
%signal-to-noise.  If the closest approach is only 10$\theta_E$, the
%interior image is $10^4$ times fainter.  For the interference pattern,
%this is a $10^{-2}$ intensity modulation since voltages are added, not
%power.  A S/N of $\gtrsim 100$ would still allow a detection.

PSRJ1745-2900 is currently about 1 million Scharzschild radii away
from the galactic center black hole, in projection. If its orbit is
inclined within 0.1\% to the line of sight, gravitational lensing
effects become order unity. In this particular system, plasma
scattering decoheres the radiation by $\delta t_s \sim ({\rm
  GHz}/\nu)^4$ sec, and at frequencies of $\sim$ THz the scattering
becomes negligible.  It is not known how bright the pulsar is at THz
frequencies.  The 
ALMA\footnote{www.almaobservatory.org} telescope might detect the
pulsar at these frequencies.  The gravitationally lensed images will
form an interference pattern if the emission size is less than $\sim
\lambda \theta_E/\theta_s$, or about 1m.  Some pulsar emission, for
example Crab giant pulses, are thought to come from regions
sufficiently compact to scintillate. Current plasma lensing limits the
emission size to be less than about a kilometer
\cite{2013arXiv1301.7505P,2012ApJ...758....8J}.  There is about a
0.1\% chance that this pulsar will have a favourable orientation to
test quantum gravity.  It would seem prudent to search for more
black-hole pulsar candidates, encouraged by the discovery of this
first one.

One might worry that plasma effects confuse quantum gravity.  While
the presence of plasma lensing could certainly overwhelm gravitational
lensing in some systems, this should not lead to a misidentification:
plasma dispersion depends on wavelength squared, while gravitational
lensing is achromatic.  Should one find a strong wavelength dependence
of the effect, one would need to find a new and cleaner system with
less plasma.  It seems unlikely to cause a gravitational
misinterpretation.  Holography of propagation effects can be used to
remove plasma distortions, or even be used constructively to resolve
the gravitationally lensed images, enabling a direct measure of phase
coherence\cite{2013arXiv1301.7505P,2013arXiv1302.1897P}.

To summarize the observational test of pulsar lensing phase coherence:
According to firewall supporters, the BHC picture predicts that the
interior image undergoes substantial phase changes, of order $r_s/b$.
For wavelength longer than that, the interference pattern should be
obervable, and for shorter wavelengths, it should weaken and
disappear.  In the course of the orbit, this will change, with the
fractional BHC impact minimized during conjunction (assuming
conjunction is outside the einstein radius).  Finite emission size
could also lead to an absence of an interference pattern.  This has a
different dependence on lensing geometry, so again seems unlikely to mimic
the BHC signal.

{\it Accretion Flow Flares} The Event Horizon Telescope
(EHT)\footnote{http://www.eventhorizontelescope.org/} could image
radiation emitted by the plasma near the horizon of the galactic
center black hole This flow is known to be unsteady, with flares and
other phenomena, that are strongly lensed\cite{2005MNRAS.363..353B}.
As for the pulsar case, the accretion flow and flare are classical
objects, and expected to follow the classical dynamics.  The low
energy photons emitted from the flare are low energy probes, and
subject to interaction in BHC.  Here, the effects would be large.  In
a fuzzball picture, the interior lensed image would appear extended,
i.e. fuzzy.  Normally, this fainter image, the one closer to the black
hole, is smaller due to conservation of surface brightness.  The
opposite could happen in a fuzzball picture: any lensed image would
maintain a constant minimum size $\theta_s\sim 20 \mu$" due to the
fuzziness.  For a flare smaller than this size, deviations will be
observable.

EHT experiments are unlikely to show interference between lensed
images, since accretion disk emission is generally extended.  Here,
the test would rely on localized flares, where multiple strongly
lensed images would be visible, and time profiles are observables.  

\textit{Discussion--} The lensing framework gives a simplified picture
to discuss Hawking radiation entanglement.  Instead of the typical
$\exp({10^{\sim 80}})$ microstates, we can focus on the low order
multipoles, reducing the variables to $a_1,\theta_1$.  To clarify the
situtation further, we consider a further restricted subspace:
$\phi_1=0$, with the axis chosen such that the classical interior
lensed image is at $\phi=0$.  We will further restrict $a_1 \in
\{-1,1\}$, i.e. a 2-state system $\ket{+},\ket{-}$, and consider
a pure state black hole.  
%This is now analogous to the Stern-Gerlach experiment:
%photons may be deflected up or down.  
Any incoming photon state becomes
entangled with the perturbed lensing eigenstates: $|i\rangle =
\alpha \ket{+,\uparrow} + \beta
  \ket{-,\downarrow}$, where $|\alpha|^2+|\beta|^2=1$.  It may be 
perturbed outward ($\uparrow$) or inward($\downarrow$), resulting in a
different phase delay as 
discussed above.  An ensemble of photons, as expected from pulsars or
flares, is projected one photon at a time.
This has some analogies to Stern-Gerlach
space quantization, with the distinction that here the
deflection field is quantum mechanical, perhaps like SQUID quantum
superposition experiments\cite{citeulike:3975975}.  The photon becomes
effectively entangled with the fuzzball.
For a static fuzzball, whose
eigenstates are not 
spherically symmetric, the interference pattern is affected.  When we
then add the degrees of freedom that the perturbations not discrete,
we see the persistence of quantum perturbations due to non-commutation
of propagation operators and fuzzball eigenstates.  

In any matter flow, for example an orbiting neutron star or accretion
flow, one needs to compare the rate at which particles exchange energy
with each other, compared to the differential energy shift from the
fuzzball entanglement.  For a star the outcome is clear: its large
internal degrees of freedom lead to decoherence, and they probe an
average space-time, which is Schwarzschild.  For the flare under
consideration, the fact that a lensing measurement requires the flare
to be small compared to its orbital radius, requires its
self-interaction to be larger than the differential tidal effect, and
the flare is also treated classically.  This interpretation differs
slightly from the commonly stated BHC picture about energy: a high
energy photon may still be seeing a quantum space time, while an
infalling observer with sufficient complexity to be considered
classical, will see the classical space-time.  An interesting
intermediate question would be the trajectory of a complex composite
particle, such as a proton.  Its constituent strongly interacting
gluons and quarks might each probe a different space-time, and the
full proton would see an averaged effect, closer to classical.

This analysis combines uncertain speculation from opposite opinion
camps on the nature of black hole evaporation, namely the firewall
group and the fuzzball group.  We caution that the scenario is by no
means inevitable, even in a fuzzball picture it is not clear that the
$O(1)$ variation of the space-time coefficients
This was argued by AMPS as a weakness of the
fuzzball picture.  
%In our universe, the black hole absorbs CMB photons
%at a rate much higher than the Hawking radiation rate, and we might
%expect a change in fuzzball state each time a CMB photon is absorbed.
%The fuzzball picture paints one of a small number of constructive
%examples, it by no means claims to enumerate all $e^{10^{80}}$ states
%describing a macroscopic black hole.

{\it Summary.}-- We have explored semiclassical quantum effects of
black holes.  A pulsar black hole binary provides a concrete setup
where such effects might become observable.  This proposed experiment
distinguishes between classical measurements, such as a star orbiting
a black hole, and quantum measurements, such as the interference of
two light paths bent by the gravitational field of the black hole.
The interference pattern could be changed by the quantum nature of the
black hole if the resolution of the Hawking paradox lies in the
non-purity of Hawking radiation.  The outcome is speculative in
nature, and is hoped to stimulate further investigation of
non-Einsteinian outcomes of strong lensing experiments.  We have
argued that the recent pulsar-black hole system, and likely more
future discoveries, allow us to probe new aspects of quantum gravity.
At least, until the ideal pulsar-black hole system is discovered it
provides a new sandbox to test ideas, to answer questions about
complementarity and the interaction of Hawking radiation with
space-time.

{\it Acknowledgements.}  I am grateful to Jason Gallichio, Avery
Broderick, Aephraim Steinberg, Daniel James and Aggie Branczyk for
stimulating discussions.  I thank the Tsinghua University for Advanced
Studies where this work was completed.

\newcommand{\araa}{ARA\&A}   % Annual Review of Astronomy and Astrophys.
\newcommand{\afz}{Afz}       % Astrofizica
\newcommand{\aj}{AJ}         % Astronomical Journal
\newcommand{\azh}{AZh}       % Astronomicekij Zhurnal
\newcommand{\aaa}{A\&A}      % Astronomy and Astrophysics
\newcommand{\aas}{A\&AS}     % Astronomy and Astrophys. Supplement Series
\newcommand{\aar}{A\&AR}     % Astronomy and Astrophysics Review
\newcommand{\apjs}{ApJS}     % Astrophysical Journal Supplement Series
\newcommand{\apjl}{ApJ}      % Astrophysical Journal Letters
\newcommand{\apss}{Ap\&SS}   % Astrophysics and Space Science
\newcommand{\baas}{BAAS}     % Bulletin of the American Astron. Society
\newcommand{\jaa}{JA\&A}     % Journal of Astronomy and Astrophysics
\newcommand{\mnras}{MNRAS}   % Monthly Notices of the Roy. Astron. Society
\newcommand{\pasj}{PASJ}     % Publ. of the Astron. Society of Japan
\newcommand{\pasp}{PASP}     % Publ. of the Astron. Society of the Pacific
\newcommand{\paspc}{PASPC}   % Publ. Astron. Soc. Pacific Conf. Proc.
\newcommand{\qjras}{QJRAS}   % Quart. Journal of the Royal Astron. Society
\newcommand{\sci}{Sci}       % Science
\newcommand{\sova}{SvA}      % Soviet Astronomy
\newcommand{\aap}{A\&A}

\bibliography{fuzzball}

%\begin{thebibliography}{99}
%\bibitem{mathur} S. Mathur
%\end{thebibliography} 

\end{document}